\documentstyle[aps,twocolumn,prl,graphicx]{revtex}
\begin{document}                % INITIALIZE - DONT CHANGE %
\draft
\wideabs{
\author{A. I. Lvovsky\cite{Lvovsky}}
\address{Fachbereich Physik, Universit\"at Konstanz, D-78457 Konstanz, Germany}
\author {J. Mlynek}
\address{Humboldt-Universit\"at zu Berlin, D-10099 Berlin, Germany}
\date{\today}
\title{Quantum-optical catalysis: generating ``Schr\"odinger kittens" by means of linear optics}

\maketitle

\begin{abstract}
We report preparation and characterization of coherent
superposition states $t|0\rangle+\alpha|1\rangle$ of
electromagnetic field by conditional measurements on a
beamsplitter. The state is generated in one of the beam splitter
output channels if a coherent state and a single-photon Fock state
$|1\rangle$ are present in the two input ports and a single photon
is registered in the other beam splitter output. The single photon
thus plays a role of a ``catalyst": it is explicitly present in
both the input and the output channels of the interaction yet
facilitates generation of a nonclassical state of light.
\end{abstract}
% \pacs{PACS numbers: 03.65.Wj, 03.67.Lx, 42.50.Dv}
}

\paragraph{Introduction} Beam splitter (BS) is the simplest quantum
optical device in which two incident light beams interfere to
produce two output beams. Quantum properties of beams splitters
are manifested in its ability to generate an entangled output
state with non-classical, but unentangled input \cite{Kim01}. This
feature has been exploited in a variety of fundamental
experiments, such as preparation of the Einstein-Podolsky-Rosen
\cite{Kimble92} and Greenberger-Horne-Zeilinger \cite{GHZExp}
states, discrete- \cite{TelepZ} and continuous-variable
\cite{TelepK} quantum teleportation. It is also an integral part
of many yet unimplemented theoretical proposals, such as
single-photon nonlocality \cite{Tan}, generating arbitrary quantum
states of light \cite{WelschPRA} and efficient linear quantum
computation \cite{LOQC,SimpleLOQC}.

In this Letter we investigate a curious consequence of beam
splitter's entangling property. A single-photon Fock state
$|1\rangle$ and a coherent state $|\alpha\rangle$ present in two
BS input channels generate an entangled state in its output [Fig.
1]. Performing measurements in one of the output channels causes
this entangled state to collapse projecting the quantum state in
the other (``signal") output onto a certain local ensemble. The
question we asked ourselves is as follows: what happens to this
ensemble if the measurement in the first output channel detects a
quantum state {\it identical} to the state in one of the BS
inputs, namely, the single-photon state? Contrary to what might be
expected we find the signal channel to be in a quantum state which
is very different from the original coherent state. This is a
quantum superposition state which possesses highly non-classical
properties such as non-positive-definite Wigner function and
subpoissonian photon statistics. We call the effect of such
transformation {\it quantum-optical catalysis} because the single
photon, while facilitating the conversion of the classical
(``target") coherent state into a non-classical signal ensemble,
itself remains unaffected by this interaction. We note that an
analogous effect of entanglement-facilitated catalysis has been
described theoretically, although for a somewhat different
setting, by Jonathan and Plenio in 1999 \cite{JP}.

In order to understand the physics of this phenomenon let us
assume that both the input coherent excitation $\alpha$ and the BS
transmission $t^2$ are small: $\alpha\sim t \ll 1$. In this case,
the coherent state can be approximated as
$|\alpha\rangle=|0\rangle+\alpha|1\rangle$. Suppose a
single-photon detector (SPD) registers a photon. Where could this
photon have originated from? If it comes from the coherent state,
the photon $|1\rangle$ from the Fock state input is likely to have
been reflected into the signal channel. If, on the other hand, the
photon detected by the SPD originates from the Fock state
transmitted through a BS, the quantum state in the signal channel
is with a high probability vacuum $|0\rangle$.

\begin{figure}
\begin{center}
\includegraphics[width=0.45\textwidth]{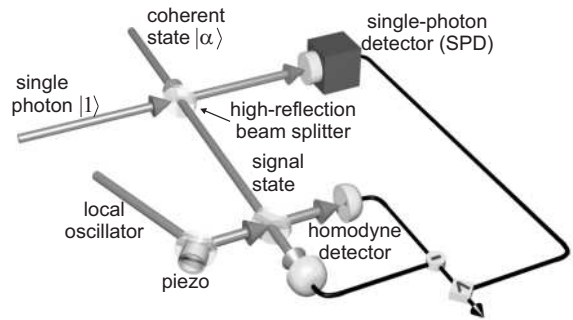}
\caption{Conceptual scheme of the experiment. Measurements by the
balanced homodyne detector are conditioned on registering a photon
by the single-photon detector.}
\end{center}
\end{figure}

The quantum properties of the beam splitter come into play when we
notice that these two possibilities are {\it fundamentally
indistinguishable}. If the two initial states are prepared in
identical optical modes, there is no way of telling which one of
the initial states the photon in the SPD channel is coming from.
As a result, the quantum state in the signal channel is {\it not a
statistical mixture of the states $|0\rangle$ and $|1\rangle$ but
their coherent superposition}, a ``Schr\"odinger kitten"
\begin{equation}
|\psi_s\rangle=\frac{1}{\sqrt{t^2+\alpha^2}}(t|0\rangle+\alpha|1\rangle).\label{psi_s}
\end{equation}

Although $|\psi_s\rangle$ has the same components as the initial
faint coherent state, their fractions can be varied randomly by
changing the ratio between $\alpha$ and $t$. For a constant, small
$t$ the increase of $\alpha$ implements a gradual transition
between highly classical ($|0\rangle$) and highly non-classical
($|1\rangle$) states of light. In this sense the state
$|\psi_s\rangle$ can be considered a bridge between the particle
and wave aspects of the electromagnetic field.

In the present paper we generate $|\psi_s\rangle$ and investigate it
by means of quantum homodyne tomography \cite{HomoTomo}. This work
was inspired in part by a recent experiment by Rarity {\it et al.}
who overlapped a single photon with a coherent state on a
beamsplitter and studied the non-classical photon statistics in the
output \cite{Rarity97}. Means of preparing the single photon in a
well-defined spatiotemporal mode and pulsed time-domain homodyne
detection have been elaborated in our group's earlier work
\cite{Fock,Shapiro,MMpaper,FockHD}. Extensive theoretical
investigation by Welsch {\it et al.} have shown that conditional
measurements on a beam splitter can generate a wide variety of
quantum states such as Schr\"odinger cats, photon-added and
photon-subtracted states \cite{WelschEPJ}. When applied in a
sequence, such measurements lead to a technique of synthesizing
random quantum states of traveling fields \cite{WelschPRA}.

Our experiment can be seen as an implementation of the first of
two stages of the non-linear sign shift quantum gate, the basic
element of the recent linear-optical quantum computation proposal
\cite{LOQC,SimpleLOQC}. Such a gate is another example of quantum
optical catalysis, because the required modification of the target
state is achieved with the two ancilla channels involved remaining
in their original quantum states.

\paragraph{Theory} We start by assuming that the input state $|1\rangle$ is prepared
with a 100-\% efficiency in an optical mode that perfectly matches
that of the coherent state
$|\alpha\rangle=e^{-|\alpha|^2/2}\sum_n\alpha^n/\sqrt{n!}|n\rangle$
at the BS input. The entangled state in the beam splitter output
can be found by applying the BS transformation operator
\begin{equation} |\Psi_{\rm
out}\rangle=\hat B|1,\alpha\rangle,
\end{equation}
where
\begin{eqnarray}
\hat B&&|m,n\rangle=\sum_{j,k=0}^{m,n}
\sqrt{\frac{(j+k)!(m+n-j-k)!}{m!\,n!}}
\left({\matrix{{m}\cr{j}\cr}}\right)
\left({\matrix{{n}\cr{k}\cr}}\right)\nonumber \\ &&\times
(-1)^{k}\, t^{n+j-k}\,r^{m-j+k}\,|j+k,\,m+n-j-k\rangle,
\end{eqnarray}
$r^2=1-t^2$ is the BS reflectivity and
$\left({\matrix{{a}\cr{b}\cr}}\right)$ are binomial coefficients
\cite{Leonhardt}.

A commercial single-photon detector generates a ``click" when
$n\ge1$ photons are incident upon its sensitive area. It is
described by the following positive operator-valued measure
(POVM):
\begin{eqnarray}
\hat\Pi_{\textrm{\scriptsize{ no-click}}}
&=&\sum_{n=0}^\infty(1-\eta_{\rm SPD})^n|n\rangle\langle
n|;\nonumber\\ \hat\Pi_{\rm click}&=&\hat
1-\hat\Pi_{\textrm{\scriptsize{ no-click}}},
\end{eqnarray}
where $\eta_{\rm SPD}$ is its quantum efficiency. A measurement
via the SPD leads to a collapse of the state $|\Psi_{\rm
out}\rangle$ projecting it, in the event of a ``click", upon the
following ensemble in the signal channel:
\begin{equation}
\hat\rho_s=\rm{Tr}_1\left(|\Psi_{\rm out}\rangle\langle\Psi_{\rm
out}|\hat\Pi_{\rm click}\right),
\end{equation}
which is readily calculated.

In our experiment the initial single photon is prepared by means
of conditional measurements on a biphoton generated via parametric
down-conversion. As discussed in our previous publications
\cite{Fock,Shapiro,MMpaper}, inefficiencies associated with this
technique result in the photon being prepared with a substantial
admixture of the vacuum state:
$\hat\rho_{|1\rangle}=\eta_{|1\rangle}|1\rangle\langle
1|+(1-\eta_{|1\rangle})|0\rangle\langle 0|$, where
$\eta_{|1\rangle}$ is the preparation efficiency. The vacuum
fraction of $\hat\rho_{|1\rangle}$, interacting on the beam
splitter with the coherent state, does not entangle itself with it
\cite{Kim01} but produces {\it coherent states} with amplitudes
$\alpha t$ and $\alpha r$ in the two BS output channels. Taking
this circumstance into account, we write the output ensemble as
\begin{equation} \hat\rho'_s=\eta'_{|1\rangle}\hat\rho_c
+(1-\eta'_{|1\rangle})|\alpha t\rangle\langle \alpha t|.
\end{equation}
Note that the quantity $\eta'_{|1\rangle}$ in the above equation
is always higher than $\eta_{|1\rangle}$ because a photon present
in the input increases the likelihood for the SPD to trigger. For
$\alpha\ll t$, $\eta'_{|1\rangle}$ tends to 1, for $\alpha\gg t$
it approaches $\eta_{|1\rangle}$.

Next we need to account for the SPD dark count events. In such an
event, the quantum state in the signal channel is not conditioned
on that in the SPD channel. The BS acts upon the incident
single-photon state simply as a lossy reflector, reducing its
efficiency by a factor of $r^2$. In addition, the coherent field
causes the phase-space displacement of this state, producing a
statistical mixture of a {\it displaced Fock state}
\cite{OKKB,DFS} and a coherent state:
\begin{equation} \hat\rho_{\rm DC}=\eta_{|1\rangle}r^2 \hat D(\alpha t)|1\rangle\langle
1|\hat D^\dagger(\alpha t) +(1-\eta_{|1\rangle}r^2)|\alpha
t\rangle\langle \alpha t|, \label{EqDFS}
\end{equation}
where $\hat D(\alpha t)$ is the displacement operator. The above
state admixes to $\hat\rho'_s$ in the proportion determined by the
SPD dark count rate.

Finally, in order to compare our prediction with the experimental
results we need to account for the non-unitary quantum efficiency
$\eta_{\rm HD}$ of the balanced homodyne detector (HD). This is
done by means of the generalized Bernoulli transformation well
described in the literature \cite{Leonhardt,Bernoulli}.

\paragraph{Experimental apparatus} The core of our apparatus consisted of
the setup for generating the single-photon Fock state, which was
the same as in our previous experiments
\cite{Fock,Shapiro,MMpaper}. A 82-MHz repetition rate train of
1.6-ps pulses generated by a Spectra-Physics Ti:Sapphire laser at
790 nm was frequency doubled and directed into a beta-barium
borate crystal for down-conversion. The down-conversion occurred
in a type-one frequency-degenerate, but spatially non-degenerate
configuration. A single-photon detector, placed into one of the
emission channels, detected photon-pair creation events. This
detector firing ensured that a photon has as well been emitted
into the other down-conversion channel, thus preparing
single-photon Fock states by conditional measurements. All further
measurements were conditioned upon a biphoton production event. We
have obtained between 300 and 400 such events in a second.

Pulses containing the conditionally-prepared photon entered an
optical arrangement shown in Fig.\,1, which had to be maintained
interferometrically stable throughout the experimental run. We
used a Perkin-Elmer SPCM-AQ-131 single-photon counting module with
$\eta_{\rm SPD}\approx0.5$ and a beam splitter with a reflectivity
of $r^2=0.92$. The choice of the BS was dictated, on one hand, by
the requirement that $t$ be small compared to 1; on the other
hand, the SPD count rate, proportional to $t^2+\alpha^2$ for small
$\alpha$, should have been sufficiently high in order to prevent
domination of the dark counts and to allow acquisition of a large
amount of data sufficient for quantum tomography.

The local oscillator for balanced homodyne detection
\cite{FockHD}, as well as the target coherent state, have been
provided by the master Ti:Sapphire laser. These pulses had to be
spatially and temporarily mode matched with each other, as well as
with the single-photon pulse emerging from the down-converter. To
this end, we first mode matched the single-photon pulse with the
local oscillator via the technique described in
\cite{Fock,MMpaper}. The coherent state was then mode matched with
the local oscillator by optimizing the visibility of the
interference pattern observed between these two classical fields.
The amplitude $\alpha$ of the target coherent state could be
varied via a variable attenuator consisting of a half-wave plate
and a polarizer.

A set of preliminary measurements was performed in order to test
the system and evaluate its main characteristics. First, we
calibrated the homodyne detector by observing the quantum noise of
the vacuum state with the beam paths of both the coherent state
and the Fock state blocked [Fig.\,2(a)]. Second, we unblocked the
coherent state input so the quantum state entering the HD was also
a coherent state with amplitude $\alpha t$ [Fig.\,2(b)].
Performing homodyne tomography on this state allowed us to
determine $\alpha$ to within $0.15$. Third, we blocked the
coherent state but unblocked the Fock state. In this case, the
quantum noise measured by the HD was associated with the
statistical mixture of the states $|1\rangle$ and $|0\rangle$ with
an efficiency $\eta_{\rm tot}=r^2\eta_{|1\rangle}\eta_{\rm HD}$
[Fig. 2(e)]. By analyzing the statistics of this noise, we found
$\eta_{\rm tot}=0.58\pm0.02$. Since it was known from previous
measurements that $\eta_{\rm HD}=0.91$ \cite{FockHD}, we estimated
$\eta_{|1\rangle}=0.69$ [Fig.\,2(c)].

Finally, we unblocked both the coherent state and the Fock state
BS inputs, and performed homodyne measurements conditioned upon
the SPD firing. The local oscillator phase was scanned by applying
a linear displacement to a mirror via a piezoelectric transducer.
The acquired data was then scaled in accordance with the vacuum
state noise acquired earlier and each data sample was associated
with the appropriate local oscillator phase value. The Wigner
function was reconstructed for each data set by means of the
inverse Radon transformation implemented via the standard filtered
back-projection algorithm \cite{Leonhardt} with a cutoff frequency
of $6.4$. This algorithm was applied directly to the quadrature
data without calculating marginal distributions associated with
each phase value \cite{DFS}.

\begin{figure}
\begin{center}
\includegraphics[width=0.45\textwidth]{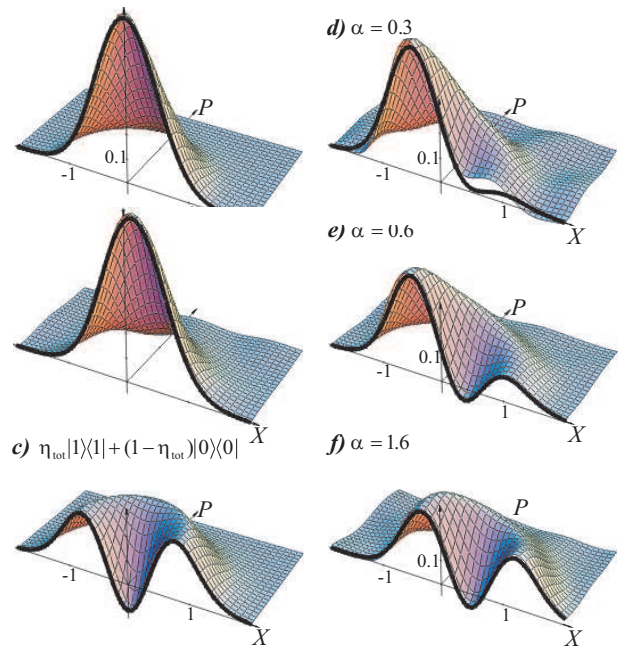}
\caption{Wigner functions of the quantum states observed in the
experiment.  (a--c): States obtained during the alignment
procedure: (a) vacuum state; (b) coherent state; (c) statistical
mixture of states $|1\rangle$ and $|0\rangle$, efficiency
$\eta_{\rm tot}=0.59$. (d--f): Quantum catalysis ensembles
obtained for various values of $\alpha$. For low $\alpha$, the
Wigner function resembles that of the vacuum state; For high
$\alpha$ -- the displaced Fock state. Heavy lines show theoretical
fits to Wigner functions' cross-section.}
\end{center}
\end{figure}

\paragraph{Results and discussion} Fig.\,2(d--f) shows the Wigner functions reconstructed for various
coherent state amplitudes. The three Wigner functions shown allow
to trace the trend the catalysis ensembles follow with increasing
$\alpha$. For $\alpha=0$, a photon registered in the SPD channel
means there is nothing in the signal channel (unless it is a dark
count); the signal state must thus be vacuum. We have not
reproduced this case experimentally due to an extremely low count
rate associated with it. With $\alpha$ increasing to the order of
$t$, the quantum state approximates the ``Schr\"odinger kitten"
(\ref{psi_s}) with a significant admixture of the vacuum due to a
non-unitary $\eta_{|1\rangle}$ [Fig.\,2(d)].

%Non-ideal properties of the SPD do not play any role in this
%regime because the actual number of photons incident rarely
%exceeds 1.

In the regime of high $\alpha$'s the SPD will fire with almost
every laser pulse detecting a large fraction of the reflecting
coherent state's photons. This mode is analogous to the dark count
case discussed earlier: the measurement of the signal state is
effectively no longer conditioned on that in the SPD channel, so a
statistical mixture (\ref{EqDFS}) of a displaced Fock state and a
coherent state obtains. The ensemble shown in Fig.\,2(f)
approaches this situation. An experimental study of displaced Fock
states will be reported separately \cite{DFS}.

The fits shown in Fig.\,2(d--f) were calculated using the
experimental parameters determined during the preliminary
measurements and varied within their tolerances in order to obtain
the best fit. An exception is made for $\eta_{\rm HD}$ which was
set to $0.83$ in order to account for the imperfect mode matching
between the coherent state and the local oscillator. Precise
treatment of this imperfection is rather complicated; reducing the
value of $\eta_{\rm HD}$ is a simplified solution which however
allows to generate an excellent fit shown in Fig.\,2(e,f). A
deviation from the perfect fit seen in Fig.\,2(d) is a technical
artefact caused by a very low data acquisition rate.

Density matrices of the catalysis ensembles have been determined
by applying the quantum state sampling method
\cite{Leonhardt,patterns} directly to the homodyne data. The
ensemble shown in  Fig.\,3 approximates a statistical mixture of
the state (\ref{psi_s}) and the vacuum. Presence of non-diagonal
elements $\rho_{01}$ and $\rho_{10}$ distinguishes the ensemble
from a statistical mixture of the states $|1\rangle$ and
$|0\rangle$ investigated in our earlier papers
\cite{Fock,Shapiro}. On the other hand, almost vanishing matrix
elements associated with states $|2\rangle$ and above shows the
difference between this density matrix and that of the coherent
state.

\begin{figure}
\begin{center}
\includegraphics[width=0.45\textwidth]{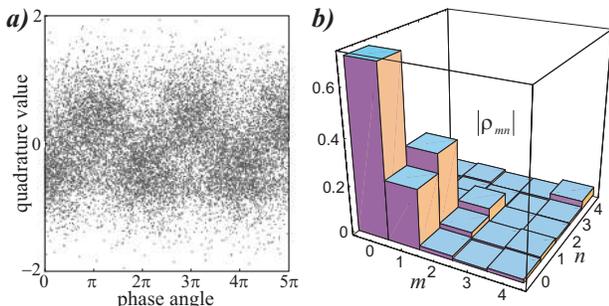}
\caption{14153 raw homodyne detector data (a) and the absolute
values of the density matrix elements in the Fock representation
(b) for the ``Schr\"odinger kitten" state with $\alpha=0.3$.}
\end{center}
\end{figure}

In conclusion, we have synthesized and characterized a new quantum
state of light, a coherent superposition of the single-photon and
vacuum states. This state was obtained by means of linear optics,
single-photon sources and detectors. This state is generated by
means of a ``catalytical" process in which the single-photon state
enabling the interaction without being affected by it. By changing
the amplitude of the target coherent state, a gradual transition
between a highly classical and a highly non-classical state of
light occurs. This experiment can be viewed as a step towards
experimental implementation of linear optical quantum computation.

This work has been sponsored by the Deutsche
Forschungsgemeinschaft and the Optik-Zentrum Konstanz. We thank S.
Babichev for experimental assistance.

%As mentioned above, in order for the ensemble $\hat\rho_c$ to
%approximate the desired state (\ref{psi_s}) we need to reduce both
%$t$ and $\alpha$. This is achieved at a cost of lowering the SPD
%count rate, which can be evaluated as $t^2+\alpha^2$.

%Always someone at your back only waiting to attack.

\end{document}